\documentclass[12pt]{amsart}
\usepackage{geometry} % see geometry.pdf on how to lay out the page. There's lots.
\usepackage{csquotes}
\title{Cosmological Black Hole Observations and Kant's Transcendental Aesthetic}
\author{Dr. Juna A. Kollmeier}
\date{} % delete this line to display the current date

\begin{document}

\maketitle

\begin{abstract}
Kant put forward the notion of the Transcendental Aesthetic (TA) in his manuscript {\it A Critique of Pure Reason}.  In this note I review the TA in light of the detection of gravitational wave radiation.  While the notion of the TA has been refuted in many different ways since its introduction, I argue that this simple proof by contradiction is of interest pedagogically and philosophically.   I hope this elucidation may be useful in introductory science courses and in science communication more generally.

\end{abstract}

\section{Introduction}
Immanuel Kant's {\it A Critique of Pure Reason} (Kant, 1781; CPR)  presents a very clear description of space and time --- one that was controversial in its day and all days after.  In this article I review the basic position held by Kant, specifically the key conceptual idea of the transcendental aesthetic (TA).  This notion has been rejected historically, most importantly in the early 20th century with the development of the mathematical framework of General Relativity which describes space and time in a completely different way from Kant, and indeed, in a material and, thus very ``real" form.   Most objectionable in the TA, from the modern perspective of Science, is that it renders us, by construction, impotent to interrogate ``reality" because it denies objects of objective reality outside of our sensory\footnote{In this discussion, we will grant Kant the vastly expanded sensory capabilities we now have thanks to progress in science and engineering such as X-ray telescopes and atomic imaging, of which he could not possibly have been aware. For the present argument these will all be considered as part of our ``experience" even though they were certainly not part of Kant's.  This prevents us from arguing, trivially, that new technology alone renders the TA null, because we will consider those advances as part of sensory experience.}  capacity.  This is not to be confused with basic empiricism and Kant himself makes this distinction.  In Kant's CPR it is not our sensory limitations that limit our knowledge of reality but rather, it is the nature of reality itself that is limited.

Supporting evidence for GR abounds and this note is not meant to deny those observations their proper credit in illuminating the nature of space, time and compact objects.  I will simply show here that the very specific claims made in CPR regarding the TA are refuted cleanly and crisply by the recent detection of gravitational wave radiation.  

\section{Proof By Contradiction}
\subsection{What Is The TA}
I first synthesize Kant's TA (via excepts) for the reader who is unfamiliar with this concept (which forms the backbone of CPR).  In the briefest of terms, the TA holds that ``space and time and all portions thereof are intuitions" (CPR pp. 486).  Kant argues: 
\begin{displayquote}
 ``Time is, therefore, merely a subjective condition of our (human) intuition (which is always sensuous), and in itself, independently of the mind or subject, is nothing." 
 \end{displayquote}
 According to Kant's TA, entities in space and time therefore do not have properties objective and unto themselves but obtain these only as organized by the consciousness which necessarily relies on sensory information.  He calls this organization the ``manifold of appearances" or, in some places in the manuscript, representations.

\smallskip
\noindent
Kant writes in the preface to the 2nd Edition in 1787:
\begin{displayquote}
 
``However harmless idealism may be considered -- although in reality it is not so --- in regard to the essential ends of metaphysics, it must still remain a scandal to philosophy and to the general human reason to be obliged to assume, as an article of mere belief, the existence of things external to ourselves (from which, yet we derive the whole material of cognition for the internal sense), and not to be able to oppose a satisfactory proof to anyone who may call it in question". 
\end{displayquote}
\smallskip
\noindent
\noindent
Kant considers it {\it scandalous} to the thinkers of his time that there should be an assumption of external and material reality without satisfactory proof.    In particular, the scandal is the {\it belief without proof} in an external reality, specifically of space and time and objects in space and time.  Having such a belief at the heart of understanding the world was intolerable for Kant.  His TA is his attempt to lift that degeneracy such that it would not infect logical inference about the world.  To serve this goal, he further holds that space is immovable and a non-entity unto itself as he writes in his chapter on direct refutation of idealism:
\begin{displayquote}
\smallskip
``The dogmatical theory of idealism is unavoidable, if we regard space as a property of things in themselves; for in that case it is, with all to which it serves as condition, a nonentity."     
\end{displayquote}
He continues:
\begin{displayquote}
\smallskip
``Space, considered in itself contains nothing movable".
\end{displayquote}
Kant thus puts forward an alternate vision of reality from that of the scientists of his time and certainly, in reflection, wholly in opposition to the modern scientific view of reality.  It is curious that this perspective seems to be gaining popularity among the public perhaps because we take the modern scientific view (that has prevailed for over 100 years) for granted and sacrosanct.
\noindent
\subsection{Refutation}
In the 21st century, we do not ``merely believe" in the existence of space-time as a movable and material entity.   GR has many successes and it was adopted rapidly because it made sense of anomalies that astronomers and physicists had struggled with (for example, the orbit of Mercury).   Even today, we do not regard GR as the ``End of Physics" --- it is the express purpose of certain theoretical branches of physics to move beyond GR and to seek deviations from GR.  However, the observation of GW150914 by the Laser Interferometry Gravitational Observatory (LIGO) demonstrates that {\it the motion of space itself} can indeed be detected.  The significance of this detection cannot be overstated.  It is not merely a ``confirmation of GR" but a confirmation that space-time and black holes are {\it real}.  As a result, just like Kant's rainbows\footnote{``Thus, we call the rainbow a mere appearance of phenomenon in a sunny shower, and the rain, the reality or thing in itself"} which he describes (incorrectly, based on incomplete understanding of light and optics) the {\it fundamental premise} of the TA is proven false by contradiction.  This contradiction comes in two forms:  both mathematical prediction (which has been the longstanding implicit refutation of the TA although rarely explicitly stated as such) and now subsequent {\it direct observation} by the LIGO experiment.   The first is a necessary condition for refutation but the latter is, in itself, sufficient. {\it Quod Erat Demonstrandum}.

\section{Context}
Having simply and rapidly dispensed with the TA, I present some historical context.  CPR was published almost exactly 100 years after Newton's {\it Principia}.  Of particular interest is that it is very nearly contemporaneous (only published 3 years prior) to the Rev. John Michell's article which is the first known instance of the term ``black hole"\footnote{Michell seems to have been enamored with Newton's gravity as he also presented a technique to measure the gravitational constant $G$, as well as the notion of plate tectonics.}.  Interestingly, it is clear upon further examination of CPR that Kant struggled mightily with the relationship between observation and theory.  In Chapter II he writes:
\begin{displayquote}
``Thoughts without content are void; intuitions without conceptions, blind.  Hence it is as necessary for the mind to make its conceptions sensuous (that is, to join to them the object in intuition), as to make its intuitions intelligible (that is, to bring them under conceptions)...Understanding cannot intuit and the sensuous faculty cannot think."
\end{displayquote}

Is this the case?  Influenced heavily by the observations of binary stars made by Sir William Herschel, Michell first invented the notion of a black hole purely from a Newtonian perspective.  By this time the now well-known formula for the escape velocity ($\sqrt{2GM/r}$, where $G$ is the grativational constant, $M$ and $r$, are the mass and radius of an object respectively) was being considered in light of celestial observations.  In his paper he argued simply (Michell, J. 1784):
\begin{displayquote}
``If there should really exist in nature any bodies, whose density is not less than that of the sun, and whose diameters are more than 500 times the diameter of the sun, since their light could not arrive at us; or if there should exist any other bodies of somewhat smaller size, which are not naturally luminous; of the existence of bodies under either of these circumstances, we could have no information from sight; yet, if any other luminous bodies should happen to revolve about them we might still perhaps from the motions of these revolving bodies infer the existence of the central ones with some degree of probability..."
\end{displayquote}

Laplace would discover something very similar ten years later, in his article ``Exposition du System du Monde" writing:
 \begin{displayquote}
``A star of the same density of the earth and with diameter 250 times larger than that of the sun would not allow, by virtue of it's gravity, even one ray of light to arrive to us."
 \end{displayquote}
 
 Of course, both Michell and Laplace argue purely from the then known Newtonian laws of motion such as the Earth, Sun and other stars which one will immediately agree are indeed objects about which we have learned from our sensory experience.  In the case of Michell it was the direct observation of binary stars that led him to consider ``escape".  However, we must also agree that they indeed intuited these objects based on their physical understanding of Newton's laws and further that they exercised this intuition in regimes well outside of their sensory capacity at the time.  Thought, content, intuition, and conception are inextricably linked in this case. It was not known whether such an object could possibly exist in the universe but the prediction is extremely clear\footnote{A. Kravtsov notes that an interpretation of ``intuition" as equivalent to ``prediction" would render Michell's black holes ``protected" by Kantian analysis.  This needs further investigation as it is not clear whether this equivalence is appropriate.}. Kant argues in the Transcendental Logic that theory is primarily ``clean up" of empirical observations.  But Michell and Laplace are at the bleeding edge of science as fundamentally {\it predictive}.  The purely mathematical construct of black holes from GR and the prediction of gravitational wave radiation bring this practice, specifically vis a vis black holes, to a qualitatively and radically new level conceptually.   
 
GR and the near-immediate ``null solution" of Schwarzschild were not derived until over 130 years after CPR and the black holes envisaged by Michell and Laplace.  The ``black hole" in this case is a mathematical singularity -- an object without a formal surface or a size as we would understand it from our sensory experience (so rather different, in character, from the black holes of the 18th century although unified, more or less, at the event horizon).  These objects are not merely mildly on the fringe of our experience, intuition and sensory capability.  They are radically so.  One must then argue that Kant's views on logic are also fundamentally flawed --- understanding {\it can} intuit.  Schwartzschild and Einstein could not have understood black holes from their experience and these black holes would not be discovered and recognized for many decades after the 1916 solution.  This is the alternative (to Kant's view) but fundamental pillar upon which modern science rests and which is too often completely obscured by operational notions of the ``Scientific Method".  Ironically, the notion of hypothesis testing, is rather along the lines that Kant contemplates and actually quite different from what scientists (or at least scientists at the forefront) actually do. 
 
\section{Why Should You Care?}

I write this for three reasons:  1) Pedagogy:  I think it is important that physicists understand, crisply and not vaguely, the philosophical significance of their work against the larger tableau of human history particularly as the practice of science is generally under attack in many forms;  2) Timeliness:  while GR has been confirmed many times before, the LIGO observations provide direct and crystal clear refutation of the Transcendental Aesthetic and thus this particular proof could not have been put forward prior to 2015\footnote{CPR contains multiple analogies that fatally suffer from the 18th century understanding (or lack thereof) of specific phenomena.  The goal here is to extract the essence of the Kantian argument -- an argument that a great number of non-scientists may indeed hold deeply despite its refutation among experts due, presumably, to poor education.};  3) Perspective:  Thomas Kuhn's groundbreaking work ``Structure of Scientific Revolutions" shows that it is of broader social interest to understand and assess the current status of our thinking as a field.  Kuhn writes:  ``To my complete surprise, that exposure to out-of-date scientific theory and practice radically undermined some of my basic conceptions about the nature of science and the reasons for its special success".   I thus find it of interest to consider the metaphysical arguments of old in order to put modern discoveries and puzzles into context. 

In the era of gravitational wave observatories, we may thus consider the scandal of which Kant was so concerned, officially and very finally over.

\section{Acknowledgements}
I thank John-Joseph Carrasco and Andrey Kravtsov for critical review of an early draft of this note and for making important suggestions and comments that improved the manuscript.  I thank Andy Gould for sustained discussions over two decades on matters of science and philosophy and specifically, black holes and materialism.   Finally, I thank Lufthansa airlines, whose A380-800 was in such an exceptional state of disrepair that I was forced to re-read one of the two books I keep on my iPhone.

\end{document}